\documentclass[12pt]{iopart}
\usepackage{graphicx}

\begin{document}

\title[Casimir-Polder interaction between an atom and
a cylinder]{Casimir-Polder interaction between an atom and
a cylinder with application to nanosystems
}
\author{
G~L~Klimchitskaya$^{1}$,
E~V~Blagov$^2$
and V~M~Mostepanenko$^{2}$}

\address{$^1$
North-West Technical University, Millionnaya St. 5,
St.Petersburg, Russia
}
\address{$^2$
Noncommercial Partnership  ``Scientific Instruments'', Moscow, Russia.}

\begin{abstract}
Recently the Lifshitz theory of dispersion forces was extended 
for the case of an atom (molecule) interacting with a plane 
surface of a uniaxial crystal or with a long solid cylinder 
or cylindrical shell made of isotropic material or uniaxial
crystal. The obtained results are applicable to nanosystems.
In particular, we investigate the Casimir-Polder interaction
between hydrogen atoms (molecules) and multi-wall carbon
nanotubes.  
It is demonstrated that the hydrogen atoms located inside
multiwall carbon nanotubes have a lower free energy 
compared to those located outside. We also perform 
comparison studies of the interaction of hydrogen atoms 
between themselves and with multi-wall carbon nanotube.
The obtained results are important for the problem
of hydrogen storage.
\end{abstract}

\section*{}
In the last few years, carbon nanotubes and other nanostructures
attracted much attention in connection with the problem of
hydrogen storage (see Refs.~\cite{1,2} for a review).
Many important properties of carbon nanotubes are determined
by the van der Waals and Casimir-Polder interactions.
In particular, the Casimir-Polder forces acting between hydrogen
atoms or molecules and carbon nanotubes play the main role in
absorption phenomena. However, for a long time they were
practically unexplored. Recently, modern methods using
density functional theory, quantum molecular dynamics and
Monte Carlo simulations were applied to investigate the
interaction of hydrogen atoms and molecules with single-wall
nanotubes [3--6] and fullerenes \cite{7,8}. 
Multi-wall carbon nanotubes with sufficiently many layers
can be modeled by a graphite cylindrical shell of some
length $L$, external radius $R\ll L$, and thickness
$d<R$. In this case the crystal optical axis $z$ is
perpendicular to the surface of the shell formed by the hexagonal
layers of a graphite lattice and the material of the shell is
described by the dielectric permittivities
$\varepsilon_x(\omega)=\varepsilon_y(\omega)$ and
$\varepsilon_z(\omega)$.

In Ref.~\cite{9} the Lifshitz formula for the Casimir-Polder
interaction between an atom and a semispace made of isotropic
material \cite{10,10a} was generalized for the cases when
a semispace is replaced by a plate of finite thickness
made of a uniaxial crystal or by a cylindrical shell made of the
same material. In this paper we apply the obtained Lifshitz-type
formulas to describe the Casimir-Polder interaction of hydrogen
atoms and molecules with multi-wall carbon nanotubes. In the
case of atom or molecule with a dynamic polarizability
$\alpha(\omega)$ at separation $a$ from a plane surface
of a plate made of a uniaxial crystal at temperature $T$
at thermal equilibrium, the free energy of the Casimir-Polder
interaction is given by \cite{9} 
\begin{eqnarray}
&&
{\cal F}^{p}(a,T)=-\frac{k_BT}{8a^3}\left\{2\alpha(0)+
\sum\limits_{l=1}^{\infty}\alpha(i\zeta_l\omega_c)
\right.
\label{eq1} \\
&&\phantom{aa}
\times\left.
\int_{\zeta_l}^{\infty}dye^{-y}\left[(2y^2-\zeta_l^2)
r_{\|}(\zeta_l,y)+\zeta_l^2
r_{\bot}(\zeta_l,y)\right]\right\}.
\nonumber
\end{eqnarray}
\noindent
Here the crystal optical axis $z$ is perpendicular to the plane
of plates $(x,y)$, $k_B$ is the Boltzmann constant, 
$\zeta_l=4\pi k_BlTa/(\hbar c)$ are the dimensionless
Matsubara frequencies, $\omega_c=c/(2a)$ is the characteristic
frequency, and the reflection coefficients are given by
\begin{eqnarray}
&&
r_{\|}(\zeta_l,y)=\frac{{\varepsilon_{xl}\varepsilon_{zl}}y^2-
f_z^2(y,\zeta_l)}{{\varepsilon_{xl}\varepsilon_{zl}}y^2+
f_z^2(y,\zeta_l)+2\sqrt{\varepsilon_{xl}\varepsilon_{zl}}y
f_z(y,\zeta_l)\mbox{coth}\left[f_z(y,\zeta_l)d/(2a)\right]},
\nonumber \\
&& \label{eq2} \\
&&
r_{\bot}(\zeta_l,y)=\frac{f_x^2(y,\zeta_l)-y^2}{y^2+f_x^2(y,\zeta_l)+
2yf_x(y,\zeta_l)\mbox{coth}\left[f_x(y,\zeta_l)d/(2a)\right]},
\nonumber
\end{eqnarray}
\noindent
where
\begin{eqnarray}
&&
f_z^2(y,\zeta_l)=y^2+\zeta_l^2(\varepsilon_{zl}-1), 
\quad
f_x^2(y,\zeta_l)=y^2+\zeta_l^2(\varepsilon_{xl}-1),
\nonumber \\
&&
\varepsilon_{zl}\equiv\varepsilon_z(i\zeta_l\omega_c),
\qquad
\varepsilon_{xl}\equiv\varepsilon_x(i\zeta_l\omega_c).
\label{eq3}
\end{eqnarray}
\noindent
In the limiting case of $d\to\infty$ Eqs.~(\ref{eq1})--(\ref{eq3})
give us the free energy of atom-semispace interaction
${\cal F}^s(a,T)$.

The free energy of the interaction between an atom and a cylinder
shell is \cite{9}
\begin{eqnarray}
&&
{\cal F}^{c}(a,T)=-\frac{k_BT}{8a^3}
\sqrt{\frac{R}{R+a}}\left\{
\vphantom{\sum\limits_{l=1}^{\infty}\int\limits_{\zeta_l}^{\infty}}
\frac{4R+3a}{2(R+a)}\alpha(0)\right.
+\sum\limits_{l=1}^{\infty}
\alpha(i\zeta_l\omega_c)
\label{eq4} \\
&&\times\int_{\zeta_l}^{\infty}dy ye^{-y}
\left[y-\frac{a}{2(R+a)}\right]
\left.\left[
\left(2-\frac{\zeta_l^2}{y^2}\right)
r_{\|}(\zeta_l,y)+
\frac{\zeta_l^2}{y^2}
r_{\bot}(\zeta_l,y)\right]\right\}.
\nonumber
\end{eqnarray}
\noindent
This formula was obtained using the proximity force theorem
and the approximate expression for the interaction energy
of two concentric cylinders \cite{11}. It is practically
exact at $a\ll R$ and has an error of about 1\% at all
$a\leq R/2$.

\begin{figure*}[t]
\vspace*{-10cm}
\includegraphics{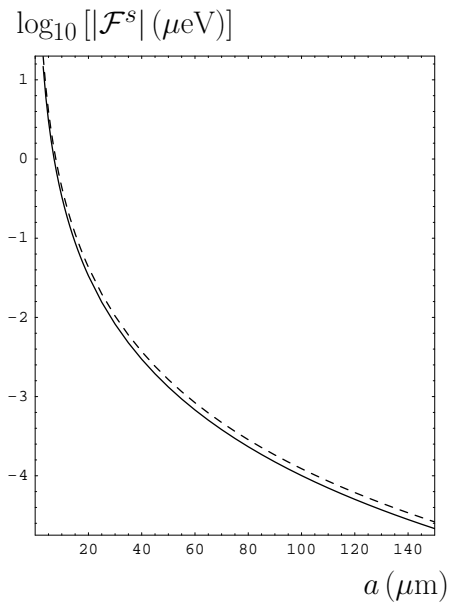}
\vspace*{-11.3cm}
\caption{
Magnitude of the free energy of a hydrogen atom (solid line) 
and molecule (dashed line) in a logarithmic scale
as a function of separation from a graphite semispace.
}
\end{figure*}
We have computed the free energy of the Casimir-Polder 
interaction between a hydrogen atom or molecule and a graphite
semispace or multi-wall carbon nanotube using Eqs.~(\ref{eq1})
and (\ref{eq4}), respectively. The dynamic polarizabilities of 
a hydrogen atom and molecule were represented in the framework 
of a single oscillator model \cite{12} 
\begin{equation}
\alpha(i\xi_l)=\alpha_{a,m}(i\xi_l)=
\frac{g_{a,m}}{\omega_{a,m}^2+\xi_l^2},
\label{eq5}
\end{equation}
\noindent
where the parameters $g_{a,m}=\alpha_{a,m}(0)\omega_{a,m}^2$
are expressed through the static polarizabilities of an
atom and a molecule $\alpha_a(0)=4.50\,$a.u.,
$\alpha_m(0)=5.439\,$a.u. and the characteristic
energies $\omega_a=11.65\,$eV and $\omega_m=14.09\,$eV,
respectively (recall that 1 atomic unit of polarizability
is equal to $1.482\times 10^{-31}\,\mbox{m}^3$).
With a precision of better than 1\% the single-oscillator model
 leads to the same results for the free energy as
the highly accurate 10-oscillator model \cite{13}. 
The dielectric permittivities of graphite
$\varepsilon_{x,z}(i\zeta_l\omega_c)$ were found by means
of dispersion relation using the tabulated optical data
for its complex index of refraction \cite{14}.
In Fig.~1 we present the computation results for the magnitudes 
of the Casimir-Polder free energy at $T=300\,$K as a function 
of separation when hydrogen atom (solid line) or molecule
(dashed line) interact with a graphite semispace. As is seen 
from Fig.~1, the magnitude of the free energy for the
hydrogen molecule is larger 
than for the atom by 33\%, 26\%, and 21.6\% at separations
$a=3\,$nm, 30\,nm, and 150\,nm, respectively.

In Fig.~2(a) we compare the free energies of hydrogen atoms
located outside and inside a multi-wall carbon nanotube.
The free energy of an atom outside a nanotube, 
${\cal F}^{c,ext}={\cal F}^{c}(a,T)$, was computed by
Eq.~(\ref{eq4}), and inside a nanotube, ${\cal F}^{c,int}$,
was obtained by the method of additive summation of the
interatomic van der Waals potentials \cite{15,16}.
In Fig.~2(a) the difference of these energies is presented
as a function of nanotube thickness in the case that both
atoms are located at a separation $a=3\,$nm from the
external and internal surfaces of a nanotube, respectively.
Nanotubes with the fixed internal radius, $R_0=10\,$nm,
are shown by the solid line, whereas nanotubes with the
fixed external radius, $R=50\,$nm, are represented by the 
dashed line. As is seen from Fig.~2(a),
${\cal F}^{c,ext}-{\cal F}^{c,int}>0$, i.e., the position
of an atom inside a nanotube is preferable. From this
figure it is also obvious that for nanotubes of fixed
thickness $d$ the value of 
${\cal F}^{c,ext}-{\cal F}^{c,int}$ is larger for the 
nanotubes of smaller external radius $R$.
Thus, a multi-wall carbon nanotube tends to accumulate the
hydrogen atoms in its interior.
 
\begin{figure}[t]
\vspace*{-7cm}
\includegraphics{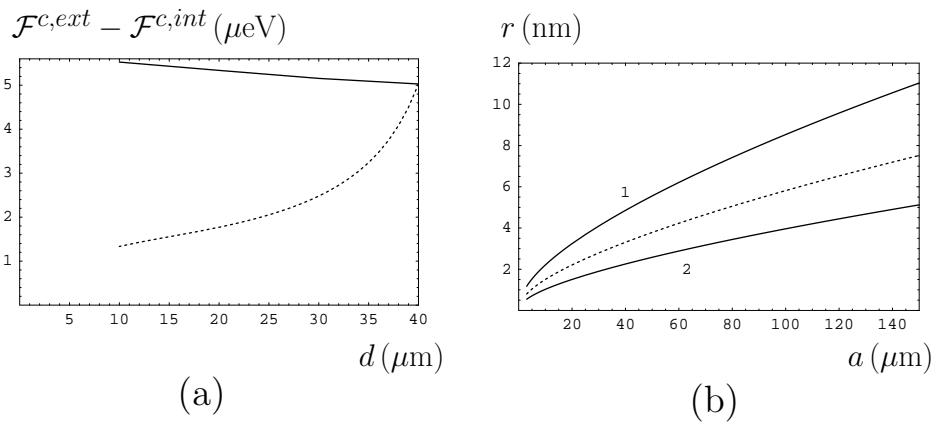}
\vspace*{-17.cm}
\caption{
(a) Difference of the free energies of two hydrogen atoms
located at $a=3\,$nm from the external and internal
surfaces of carbon nanotube with fixed internal radius
$R_0=10\,$nm (solid line) and fixed external radius
$R=50\,$nm (dashed line) as a function of nanotube
thickness. (b) Regions in the plane $(a,r)$ where the
interatomic interaction of two hydrogen atoms is by at
least 10 times smaller than their interaction to
carbon nanotube with $R=50\,$nm, $d=30\,$nm
(region above line 1), by at least 10 times larger 
(region below line 2) and where both interactions
are equal (the dashed line).
}
\end{figure}
In Fig.~2(b) we compare the van der Waals interaction
energy between two hydrogen atoms separated by a distance 
$r$ to their interaction ${\cal F}^c$ with a multi-wall 
carbon nanotube. In the region of $(a,r)$-plane above 
line 1 the interatomic interaction is smaller than the
interaction with a nanotube by at least a factor of 10.
In a similar manner, in the region below line 2 one
can neglect by the interaction with a nanotube in
comparison with the interatomic interaction. As to the
region in between lines 1 and 2, here both interactions 
are of the same order of magnitude and cannot be 
considered independently.
The figures similar to Fig~2(b) permit to choose the
conditions whereby the interatomic interactions can
be neglected in comparison with the interaction of
an atom with a nanotube.

The above results demonstrate that the multi-wall carbon 
nanotubes are of considerable promise for the problem
of hydrogen storage and deserve further investigation.

\section*{Acknowledgments}
This work was supported by the Russian Foundation for
Basic Research (Grant No. 05--08--18119a).


\section*{References}
\numrefs{99}
\bibitem{1}
Calbi M M, Cole M W, Gatica S M, Bojan M J
and Stan G 2001
{\it Rev. Mod. Phys.} {\bf 73} 857
\bibitem{2}
Ding R G, Lu G Q, Yan Z F and Wilson M A 2001
{\it J.  Nanosci. Nanotech.} {\bf 1} 7 
\bibitem{3}
Yildirim T and  Ciraci S 2005
{\it Phys. Rev. Lett.} {\bf 94} 175501 
\bibitem{4}
Bondarev I V and Lambin P 2005
{\it Phys. Rev.} B {\bf 72} 035451 
\bibitem{5}
Dobson J F and Rubio A 2005
{\it Preprint} cond-mat/0502422
\bibitem{6}
Dag S, Ozturk Y, Ciraci S and 
Yildirim T 2005
{\it Phys. Rev.} B {\bf 72} 155404
\bibitem{7}
Yildirim T, \'{I}\~{n}iguez J and  Ciraci S 2005 
{\it Phys. Rev.} B {\bf 72} 153403
\bibitem{8}
Turnbull J D and Boninsegni M 2005
{\it Phys. Rev.} B {\bf 71} 205421
\bibitem{9}
Blagov E V, Klimchitskaya G L and 
Mos\-te\-pa\-nen\-ko V M 2005
{\it Phys. Rev.} B {\bf 71} 235401
\bibitem{10}
Lifshitz E M and Pitaevskii L P 1980
{\it Statistical Physics}, Part.~II (Oxford: Pergamon Press)
\bibitem{10a}
Milonni P W 1994
{\it The Quantum Vacuum}
(San Diego: Academic Press)
\bibitem{11}
Mazzitelli F D 2004
{\it Quantum Field Theory Under the Influence of External
Conditions} (Princeton: Rinton Press, Princeton) p~126
\bibitem {12}
Rauber S, Klein J R, Cole M W and Bruch L W 1982
{\it Surf. Sci.} {\bf 123} 173 
\bibitem {13}
Johnson R E, Epstein S T and Meath W J 1967
{\it J. Chem. Phys.} {\bf 47} 1271 
\bibitem{14}
{\it Handbook of Optical Constants of Solids},
ed. Palik E D 1985 (New York: Academic Press)
\bibitem{15}
Mostepanenko V M and Trunov N N 1997
{\it The Casimir Effect and its Applications}
(Oxford: Clarendon Press)
\bibitem{16}
Bordag M, Mohideen U and Mostepanenko V M 2001
{\it Phys. Rep.} {\bf 353} 1
\endnumrefs
\end{document}